\documentstyle[12pt]{article}

\begin{document}

\title{Perturbations in {\it k-}inflation}
\author{Jaume Garriga$^{1,2}$ and V.F. Mukhanov$^{1}$}
\maketitle
\centerline{$^1$Ludwig-Maximilians-Universit\"at, Sektion Physik,} %
\centerline{M\"unchen, Germany.} 
\centerline{$^2$IFAE, Departament de F\'\i sica, Universitat Aut\`onoma
de Barcelona,} \centerline{08193 Bellaterra, Spain} \medskip
\begin{abstract}
We extend the theory of cosmological perturbations to the case when the
``matter'' Lagrangian is an arbitrary function of the scalar field and its
first derivatives. In particular, this extension provides a unified
description of known cases such as the usual scalar field and the
hydrodynamical perfect fluid. In addition, it applies to the recently
proposed $k$-inflation, which is driven by non-minimal kinetic terms in the
Lagrangian. The spectrum of quantum fluctuations for slow-roll and power law 
$k$-inflation is calculated. We find, for instance, that the usual
``consistency relation'' between the tensor spectral index and the relative
amplitude of scalar and tensor perturbations is modified. Thus, at least in
principle, {\it k-}inflation is phenomenologically distinguishable from
standard inflation.
\end{abstract}

{\it Introduction. }The quantum theory of cosmological perturbations is one
of the most interesting examples of quantum field theory in an external
classical field. Indeed, the spectrum of metric perturbations generated
during inflation is probably the only prediction of quantum field theory in
external fields (besides the Casimir effect) which can be tested
``experimentally'' in the near future. The quantum theory of linearized
cosmological perturbations is well developed in the case of a usual scalar
field with the standard kinetic term and arbitrary potential, and also for a
hydrodynamical perfect fluid \cite{rev}. This theory is sufficient to
predict the main features of the generated spectrum for a wide class of
inflationary models, as long as the accelerated expansion of the early
Universe is due to the ``potential'' energy of a scalar field. Recently,
however, a different mechanism for driving inflation, called $k-$inflation,
was presented in \cite{kinf}. In $k-$inflation, the accelerated expansion is
due to non-minimal kinetic terms in the action, and the existing theory of
cosmological perturbations does not apply. Motivated by this development, we
extend here the quantum theory of cosmological perturbation to the case of
the most general local Lagrangian which depends on the scalar field and its
first derivatives. This extension is actually quite versatile. It allows us
to describe in a unified way the ``old'' cases of the usual scalar field and
hydrodynamics, and it is also applicable to the ``new '' models admitting
kinetic inflation. As we shall see, some of the basic predictions for the
cosmological perturbation spectrum in {\it k}-inflation can be quite
different from the ones in standard inflation. In this paper, we shall
concentrate in the derivation of the main formal results. A detailed
analysis of the perturbation spectra in phenomenologically viable models 
{\it k}-inflation will be the subject of a forthcoming publication\cite{phen}%
.

{\it Setup. }We consider the most general local action for a scalar field
coupled to Einstein gravity, which involves at most first derivatives of the
field, 
\begin{equation}
S=-{\frac{1}{16\pi G}}\int \sqrt{g}Rd^{4}x+\int \sqrt{g}\ p(X,\varphi
)d^{4}x,  \label{lagrangian}
\end{equation}
where $g$ is the determinant of the metric, $R$ is the Ricci scalar and 
\begin{equation}
X={\frac{1}{2}}g^{\mu \nu }\partial _{\mu }\varphi \partial _{\mu }\varphi .
\label{defX}
\end{equation}
The Lagrangian for the scalar field is called here $p$ because it plays the
role of pressure \cite{kinf}. Indeed, varying the matter Lagrangian with
respect to the metric we obtain the energy momentum tensor in the form 
\begin{equation}
T_{\mu \nu }=(\varepsilon +p)u_{\mu }u_{\nu }-pg_{\mu \nu }.  \label{EMT}
\end{equation}
where 
\[
u_{\mu }={\frac{\varphi ,_{\mu }}{(2X)^{1/2}}} 
\]
and the energy density is given by 
\begin{equation}
\varepsilon =2X\ p_{,X}-p.  \label{energy}
\end{equation}
Here $p_{,X}$ denotes partial derivative of the Lagrangian with respect to $%
X $.

We see from (\ref{EMT}) that the Lagrangian (\ref{lagrangian}) can be used
to describe the potential motions of hydrodynamical fluid as well as to draw
a useful analogy with hydrodynamics in the case of arbitrary Lagrangian for
a scalar field. Indeed, if $p$ depends only on $X,$ then $\varepsilon
=\varepsilon (X)$. In many cases the equation (\ref{energy}) can be solved
giving the equation of state $p=p(\varepsilon )$ for an ``isentropic''
fluid. For instance, if $p\propto X^{n},$the corresponding equation of state
is $p=\varepsilon /\left( 2n-1\right) .$ In the general case, when $%
p=p\left( X,\varphi \right) ,$ the pressure cannot be expressed only in
terms of $\varepsilon ,$ since $\varphi $ and $X$ are independent. However,
even in this case the hydrodynamical analogy is still rather useful.

{\it Background.} As a background we consider an expanding Friedmann
Universe with an arbitrary spatial curvature ($K=0,\pm 1$): 
\begin{equation}
ds^{2}=dt^{2}-a^{2}\left( t\right) \gamma _{ik}dx^{i}dx^{k},
\end{equation}
where the spatial metric $\gamma _{ik}$ can describe flat, open or closed
Universe. Two independent equations for two unknown background variables $%
\varphi _{0}\left( t\right) $ and $a\left( t\right) $ can be written down in
the familiar form 
\begin{equation}
H^{2}+\frac{K}{a^{2}}=\frac{8\pi G}{3}\varepsilon ,  \label{hubble}
\end{equation}
\begin{equation}
\dot{\varepsilon}=-3H\left( \varepsilon +p\right) ,  \label{conservationeq}
\end{equation}
where $H\equiv \dot{a}/a$ is the Hubble constant and a dot denotes
derivative with respect to the physical time $t.$ For later reference it is
also useful to write down other redundant forms of these equations.
Substituting (\ref{hubble}) in (\ref{conservationeq}) we immediately get 
\begin{equation}
\dot{H}=\frac{K}{a^{2}}-4\pi G\left( \varepsilon +p\right) .
\label{derivativeofhubble}
\end{equation}
Taking into account that $\dot{\varepsilon}=\varepsilon _{,X}\dot{X}%
+\varepsilon _{,\varphi }\dot{\varphi}$ and using (\ref{conservationeq}) we
obtain the following expression for the time derivative of the pressure 
\begin{equation}
\dot{p}=p_{,X}\dot{X}+p_{,\varphi }\dot{\varphi}=-3c_{S}^{2}H\left(
\varepsilon +p\right) +\dot{\varphi}(p_{,\varphi }-c_{S}^{2}\varepsilon
_{,\varphi }).  \label{derivativeofpressure}
\end{equation}
Here we have introduced the notation 
\begin{equation}
c_{S}^{2}\equiv \frac{p_{,X}}{\varepsilon _{,X}}=\frac{\varepsilon +p}{%
2X\varepsilon _{,X}}.  \label{speedofsound}
\end{equation}
As we will see later, $c_{S}$ plays the role of the ``speed of sound'' for
perturbations.

{\it Equations for perturbations. }Considering small inhomogeneities of the
scalar field$,$ that is 
\begin{equation}
\varphi \left( t,x\right) =\varphi _{0}\left( t\right) +\delta \varphi
\left( t,x\right) ,
\end{equation}
and taking into account that $\delta T_{k}^{i}\propto \delta _{k}^{i}$ one
can write the metric of the ``perturbed Universe'' in the longitudinal gauge 
as 
\cite{rev} 
\begin{equation}
ds^{2}=\left( 1+2\Phi \right) dt^{2}-\left( 1-2\Phi \right) a^{2}\left(
t\right) \gamma _{ik}dx^{i}dx^{k},
\end{equation}
where $\Phi $ is the newtonian gravitational potential.

The perturbations $\delta \varphi $ induce the appropriate perturbations in
the components of the energy momentum tensor (\ref{EMT}), which can be
easily expressed in terms of $\delta \varphi /\dot\varphi _{0}\left(
t\right) $ and $\Phi $ as 
\begin{equation}
\delta T_{0}^{0}=\delta \varepsilon =\varepsilon _{,X}\delta X+\varepsilon
_{,\varphi }\delta \varphi =\frac{\varepsilon +p}{c_{S}^{2}}\left[ \left( 
\frac{\delta \varphi }{\dot{\varphi}}\right) ^{.}-\Phi \right] -3H\left(
\varepsilon +p\right) \frac{\delta \varphi }{\dot{\varphi}}  \label{EMT00}
\end{equation}
and 
\begin{equation}
\delta T_{i}^{0}=\left( \varepsilon +p\right) \left( \frac{\delta \varphi }{%
\dot{\varphi}}\right) _{,i}  \label{EMT0i}
\end{equation}
In deriving (\ref{EMT00}) we used the definition (\ref{defX}) of $X$ and the
equations (\ref{conservationeq}), (\ref{speedofsound}). We also skip from
here the index 0 for the background value of $\varphi \left( t\right) .$
These expressions for $\delta T$ should be substituted into the 00 and 0i
linearized Einstein equations 
\begin{equation}
\frac{1}{a^{2}}\Delta \Phi -3H\dot{\Phi}+3\left( \frac{K}{a^{2}}%
-H^{2}\right) \Phi =4\pi G\delta T_{0}^{0},  \label{00eq}
\end{equation}
\begin{equation}
(\dot{\Phi}+H\Phi )_{,i}=4\pi G\delta T_{i}^{0}.  \label{0ieq}
\end{equation}
The other Einstein equations give no additional information. They can be
obtained as a combination of the above equations and we shall not write them
down here. The equations (\ref{00eq}-\ref{0ieq}) with $\delta T$ given above
can be cast in the following form 
\begin{equation}
\left( \frac{\delta \varphi }{\dot{\varphi}}\right) ^{.}=\left( 1+\frac{%
c_{S}^{2}\left( \Delta +3K\right) }{4\pi Ga^{2}\left( \varepsilon +p\right) }%
\right) \Phi ,  \label{eq00}
\end{equation}
\begin{equation}
\left( a\Phi \right) ^{.}=4\pi Ga\left( \varepsilon +p\right) \frac{\delta
\varphi }{\dot{\varphi}}.  \label{eq0i}
\end{equation}
A more pleasant form of these equations is obtained by changing from the
independent variables $\Phi $ and $\delta \varphi /\dot{\varphi}$ , to the
new variables $\xi $ and $\zeta $ defined via 
\begin{equation}
\Phi a=4\pi GH\xi ,  \label{def1}
\end{equation}
\begin{equation}
\frac{\delta \varphi }{\dot{\varphi}}=\frac{\zeta }{H}-\left( \frac{4\pi G}{a%
}-\frac{K}{a^{3}\left( \varepsilon +p\right) }\right) \xi .  \label{def2}
\end{equation}
Substituting (\ref{def1}-\ref{def2}) into (\ref{eq00}-\ref{eq0i}) and using
the background equations (\ref{hubble}-\ref{speedofsound}) we easily obtain
the following equations for $\xi $ and $\zeta :$%
\begin{equation}
\dot{\xi}=\frac{a\left( \varepsilon +p\right) }{H^{2}}\zeta ,  \label{pereq1}
\end{equation}
\begin{equation}
\dot{\zeta}=\frac{c_{S}^{2}H^{2}}{a^{3}\left( \varepsilon +p\right) }\left(
\Delta +Kf\right) \xi  \label{pereq2}
\end{equation}
where 
\begin{equation}
f=\frac{\dot{\varphi}\left( c_{s}^{-2}p_{,\varphi }-\varepsilon _{,\varphi
}\right) }{H\left( \varepsilon +p\right) }.  \label{deff}
\end{equation}
The linearized equations (\ref{pereq1}-\ref{pereq2}) significantly simplify
in two important cases: a) in spatially flat Universe $K=0$ (for an
arbitrary $p)$ and b) for ``hydrodynamical'' matter /or scalar field without
potential (for an arbitrary $K$ ). In both cases the term proportional $Kf$
in (\ref{pereq2}) vanishes.

{\it Action. }In order to normalize the amplitude of quantum fluctuations,
the action is needed. This can be obtained by expanding (\ref{lagrangian})
to second order in perturbations. After use of the constraints, this
expansion is reduced to an expression containing only the physical degrees
of freedom \cite{rev}. However, those steps are rather cumbersome and in
fact we do not need to pass through them. Up to an overall time independent
factor, the action for perturbations can be unambiguously inferred directly
from the equations of motion (\ref{pereq1}-\ref{pereq2}). The overall factor
can then be fixed comparing this action with previously derived actions for
hydrodynamical matter and scalar field with minimal kinetic term in $K=0$ 
\cite{rev} as well as $K\neq 0$ universes \cite{gmst}, which are particular
cases of our general matter Lagrangian. The first order action which
reproduces the equations of motion (\ref{pereq1}-\ref{pereq2}) is 
\begin{equation}
S=\int \left[ \xi \widehat{O}\dot{\zeta}-\frac{1}{2}\frac{H^{2}c_{S}^{2}}{%
a^{3}\left( \varepsilon +p\right) }\xi \widehat{O}\left( \Delta +Kf\right)
\xi +\frac{1}{2}\frac{a\left( \varepsilon +p\right) }{H^{2}}\zeta \widehat{O}%
\zeta \right] dtd^{3}x  \label{action1}
\end{equation}
where $\widehat{O}\equiv \widehat{O}\left( \Delta \right) $ is a time
independent operator which still should be determined. Expressing $\xi $ in
terms of $\dot{\zeta}$ via equation (\ref{pereq2}) and substituting it in (%
\ref{action1}) one gets 
\begin{equation}
S=\frac{1}{2}\int z^{2}\left[ \zeta^{\prime}{}^2+c_{S}^{2}\zeta 
({\Delta +Kf})\zeta \right] d\eta d^{3}x  \label{action2}
\end{equation}
Here we use the conformal time $\eta =\int dt/a$ instead of physical time $t$
and prime denotes the derivative with respect to $\eta .$ The variable $z$
is defined as 
\begin{equation}
z\equiv \frac{a\left( \varepsilon +p\right) ^{1/2}}{c_{S}H}
\left( \frac{\widehat{O}\left(
\Delta \right) }{\Delta +Kf}\right)^{1/2} .  \label{defz}
\end{equation}
Here, the Laplacian should be understood as a $c$-number, representing the
corresponding eigenvalue. Introducing the canonical quantization variable 
$v=z\zeta $ one can rewrite
the action (\ref{action2}) in terms of $v.$ Comparing it with the actions
obtained in the literature for various particular cases \cite{rev,gmst}, we
infer that $\widehat{O}=\Delta +3K.$ Therefore the final result is 
\begin{equation}
S=\frac{1}{2}\int \left[ v^{\prime }{}^2+
c_{S}^{2}v\left( \Delta +Kf\right) v+\frac{z^{\prime
\prime }}{z}v^{2}\right] d\eta d^{3}x.  \label{action3}
\end{equation}
In a flat Universe this action and the expression for $z$ 
significantly simplify. In this case the equation for the variable 
$v$ {\it \ }immediately follows from (\ref{action3}) with $K=0$, and takes 
the form 
\begin{equation}
v^{\prime \prime }-c_{S}^{2}\Delta v-\frac{z^{\prime \prime }}{z}v=0,
\label{eqv}
\end{equation}
where $c_{S}^{2}$ is defined in (\ref{speedofsound}) and $z$ in 
(\ref{defz}). This equation has the obvious long-wavelength solution $v\propto z.$

{\it Power spectra. }For the remainder of the paper, we shall restrict our
attention to the case $K=0$. We shall be interested in calculating the
spectrum of the quantity $\zeta =v/z.$ This quantity is directly related to
the gravitational potential $\Phi $ in terms of which the fluctuations of
the temperature of background radiation in large angular scales are
expressed as $\delta T/T\approx \Phi /3.$ In fact, from eqs. (\ref{eq0i}-\ref
{def2}) we immediately find that $\zeta $can be expressed in terms of the
potential $\Phi $, 
\begin{equation}
\zeta =\frac{5\varepsilon +3p}{3\left( \varepsilon +p\right) }\Phi +\frac{2}{%
3}\frac{\varepsilon }{\varepsilon +p}\frac{\dot{\Phi}}{H}
\end{equation}
Since the nondecaying long-wavelength solution of (\ref{eqv}) is
proportional to $z, $ the variable $\zeta $ remains constant in this limit
irrespectively to what happens with the equation of state. On the other
hand, the nondecaying mode of the potential $\Phi$ is constant during any
period where $p/\varepsilon $ is constant \cite{rev}. Therefore, for this
mode $\Phi =3[\left( \varepsilon +p\right) /\left( 5\varepsilon +3p\right)
]\zeta .$ It is well known how to quantize the theory with the action (\ref
{action3}) and we skip the standard steps (see, for instance \cite{rev})
going directly to the spectral density 
\begin{equation}
P_{k}^{\zeta }\equiv \frac{1}{2\pi ^{2}}\left| \zeta _{k}\right| ^{2}k^{3}=%
\frac{k^{3}}{2\pi ^{2}}\frac{\left| v_{k}\right| ^{2}}{z^{2}}.  \label{spec0}
\end{equation}
This characterizes the squared dimensionless amplitude of the perturbations
in commoving scale $\lambda =2\pi /k.$ Here $v_{k}$ is the solution of the
mode function equation 
\begin{equation}
v_{k}^{\prime \prime }+\left( c_{S}^{2}k^{2}-\frac{z^{\prime \prime }}{z}%
\right) v_{k}=0,  \label{modeeq}
\end{equation}
which follows from (\ref{eqv}) by substituting $v\propto v_{k}\left( \eta
\right) \exp (i{\bf kx),}$ with initial conditions corresponding to the
minimal quantum fluctuations of the field $v.$ When $c_{S}^{2}$ changes
adiabatically, these quantum fluctuations can be unambiguously defined for
those perturbations for which the potential term $z^{\prime \prime }/z$ is
small compared to $c_{S}^{2}k^{2}$.

During ``slow roll'' inflation, the Hubble rate $H$, as well as $c_{s}$ and $%
(\varepsilon +p)$ change much slower than the scale factor. Thus, from (\ref
{defz}) we have $z^{\prime \prime }/z\approx a^{\prime \prime }/a\approx
2(Ha)^{2}$, where we have used $(\varepsilon +p)/\varepsilon \ll 1$. For a
given wavenumber $k$, the potential term in (\ref{modeeq}) can be neglected
at sufficiently early times, when the physical wavelength of the
perturbation $a/k$ is much smaller than the ``sound'' horizon $c_{s}H^{-1}$.
In the short wavelength limit, the normalized positive frequency modes
corresponding to the minimal quantum fluctuations take the form 
\begin{equation}
v_{k}\approx {\frac{e^{-ikc_{s}\eta }}{(2c_{s}k)^{1/2}}},\quad (aH\ll c_{s}k)
\label{sw}
\end{equation}
The potential term $z^{\prime \prime }/z$ starts to dominate after these
modes cross the ``sound'' horizon. Thus the solution (\ref{sw}) evolves into
the long wavelength solution 
\begin{equation}
v_{k}\approx C_{k}z,\quad (aH\gg c_{s}k)  \label{lw}
\end{equation}
where $C_{k}$ is a constant. This constant is found by matching both
solutions in the standard way. In the transition region, the universe can be
approximated by de Sitter $a\approx (H\eta )^{-1}$ and (\ref{modeeq})
becomes a Bessel equation. With the initial conditions given by (\ref{sw})
it is readily found that $|C_{k}|^{2}=(2c_{s}kz_{s}^{2})^{-1}$, where $z_{s}$
is the value of $z$ at the moment of sound horizon crossing ( $aH=c_{s}k$).
Substituting (\ref{lw}) in (\ref{spec0}) we find that the spectrum for long
wavelength perturbations is given by 
\begin{equation}
P_{k}^{\zeta }=\frac{16}{9}\left. \left( \frac{\varepsilon }{\varepsilon
_{Pl}}\right) {\frac{1}{c_{s}(1+p/\varepsilon )}}\right| _{s},
\label{scalarp}
\end{equation}
where $\varepsilon _{Pl}=G^{-2}$ is the Planckian energy density and the
subscript $s$ means that the appropriate quantity should be calculated at
the moment defined by $aH=c_{S}k.$ This equation reduces to the standard
result for slow-roll inflation driven by a potential term ($c_{s}=1$).
However, we see that in the general case the speed of sound appears in the
denominator.

The spectrum (\ref{scalarp}) depends on the wavelength because the
parameters in the right hand side are slowly varying with time, and so they
will take different values as each co-moving wavelength crosses the sound
horizon. The scalar spectral index $n_{S}$ is equal 
\begin{equation}
n_{S}-1\equiv \frac{d\ln P_{k}^{\zeta }}{d\ln k}=-3\left( 1+\frac{p}{%
\varepsilon }\right) -\frac{1}{H}\left( \ln \left( 1+\frac{p}{\varepsilon }%
\right) \right) ^{.}-\frac{1}{H}\left( \ln c_{S}\right) ^{.}+...
\label{sindex}
\end{equation}
where dot denotes the derivative with respect to $t$ taken at the moment of
sound horizon crossing and we skipped here the terms which are higher order
in slow roll parameters $\left( 1+p/\varepsilon \right) ,$ $\left( \ln
c_{S}\right) ^{.}/H$ and their derivatives. Note that all the terms on the
right hand side of the equation (\ref{sindex}) have definite sign in generic
models of inflation. Therefore inflation doesn't generically predict a
``Harrison-Zeldovich spectrum'' for scalar perturbations, as it is quite
often mistakenly stated, but it predicts rather a tilted spectrum with $%
n_{S}<1.$ The expression (\ref{sindex}) differs from the appropriate
expression in the case of usual inflation by the term proportional to the
derivative of the speed of sound, which does not vanish in slow roll {\it k}%
-inflation \cite{kinf}. We may add therefore, that when non-minimal kinetic
terms are allowed, the possibility of locally reconstructing the Lagrangian
from the power spectra meets with new obstacles. First of all, not only the
potential term has to be reconstructed. And second, knowledge of the
spectral indices does not give as much information about the slow roll
parameters as it does in standard inflation.

The spectrum of tensor perturbations is given by the usual expression 
\begin{equation}
P_{k}^{h}=\frac{128}{3}\left. \left( \frac{\varepsilon }{\varepsilon _{Pl}}%
\right) \right| _{k=Ha},  \label{tensorp}
\end{equation}
where the appropriate quantities should be estimated at the moment horizon
crossing $k=aH$. This is not exactly the same time as the time of sound
horizon crossing, but to lowest order in the slow-roll parameters this
difference is unimportant. Therefore, the ratio of tensor to scalar
amplitudes for the power spectra is given by 
\begin{equation}
{\frac{P^{h}}{P^{\zeta }}}=24c_{s}\left( 1+\frac{p}{\varepsilon }\right) .
\label{ratio}
\end{equation}
The tensor spectral index is defined as 
\begin{equation}
n_{T}={\frac{d\ln P_{k}^{h}}{d\ln k}}\approx -3\left( 1+\frac{p}{\varepsilon 
}\right) .  \label{stensor}
\end{equation}
In standard inflation ($c_{s}=1$), this spectral index and the ratio (\ref
{ratio}) are not independent, and one has the so-called ``consistency
relation'' \cite{cons} $P^{h}/P^{\zeta }=-8n_{T}$, which in principle can be
tested against observations. However, it is clear that this relation does
not hold in $k$-inflation. Instead of that we have 
\begin{equation}
{\frac{P^{h}}{P^{\zeta }}=-8}c_{s}n_{T}  \label{consis}
\end{equation}
Thus, at least in principle, kinetic inflation is phenomenologically
distinguishable from standard inflation.

When we depart from slow roll, some extra care has to be taken in comparing
the scalar and tensor amplitudes, since as mentioned above the time of
horizon crossing is not the same as the time of sound horizon crossing. An
example is discussed in the next section.

{\it Power law k-inflation. }As shown in \cite{kinf}, a Lagrangian of the
form $p=g(X)\varphi ^{-2}$, where $g$ is an arbitrary function, can drive a
power law expansion of the form 
\begin{equation}
a\propto t^{\beta }.  \label{plaw}
\end{equation}
The solutions are characterized by $X=X_{0}=$const. where $X_{0}$ is a
solution of the equation 
\[
\left( {6\pi G}g_{,X}^{2}-g_{,X}+{\frac{g}{2X}}\right) _{X=X_{0}}=0. 
\]
The power exponent is given then by $\beta ={4\pi Gg_{,X}}\left(
X_{0}\right) $. During power law inflation, the speed of sound is a constant 
$c_{s}^{2}=$$g_{,X}/\left( g{_{,X}}+{2Xg_{,XX}}\right) $. This constant is
unrelated to the power exponent, and so we can think of $\beta $ and $c_{s}$
as parameters which can be freely adjusted by taking an appropriate function 
$g.$ In particular, one can easily write down a Lagrangian for which $c_{S}>1
$. This interesting possibility will be discussed in more detail later.

It is clear from (\ref{defz}) that $z$ is just proportional to the scale
factor $a$. Reexpressing the scale factor in terms of conformal time $\eta $
and substituting it in (\ref{modeeq}) we obtain the Bessel equation for $%
v_{k}$ 
\begin{equation}
v_{k}^{\prime \prime }+\left[ c_{s}^{2}k^{2}-\left( \nu ^{2}-{\frac{1}{4}}%
\right) \frac{1}{\eta ^{2}}\right] v_{k}=0,
\end{equation}
where $\nu =(3/2)+(\beta -1)^{-1}.$ In what follows, we shall restrict
attention to inflationary solutions $\beta >1$.

Up to an irrelevant constant phase, the ``positive frequency'' solution $%
v_{k}=(1/2)(-\pi \eta )^{1/2}H_{\nu }^{(1)}(-kc_{s}\eta )$ reduces to (\ref
{sw}) in the short wavelength limit ($-kc_{s}\eta \gg 1$). At late times ($%
-kc_{s}\eta \ll 1$) this solution behaves as 
\[
{\frac{v_{k}}{z}}\approx {\frac{2^{\nu -{\frac{3}{2}}}\Gamma (\nu )}{%
(2kc_{s})^{1/2}\Gamma (3/2)}}\ {\frac{(-kc_{s}\eta )^{{\frac{1}{2}}-\nu }}{%
z(\eta )}}. 
\]
The right hand side of this equation is independent of time, and so we can
evaluate it at any convenient time. It is customary to evaluate it at the
time $\eta _{1}$ when a fiducial commoving wavenumber $k_{1}$ crosses the 
{\em cosmological} horizon, $k_{1}=a_{1}H_{1}=-\beta \lbrack (\beta -1)\eta
_{1}]^{-1}.$ Substituting in (\ref{spec0}) we have 
\begin{equation}
{P}_{k}^{\zeta }=\left[ 2^{\frac{2}{\beta -1}}\left| {\frac{\Gamma (\nu )}{%
\Gamma (3/2)}}\right| ^{2}\left( {\frac{\beta -1}{\beta }}\right) ^{\frac{%
2\beta }{\beta -1}}\right] {\frac{GH_{1}^{2}}{\pi }}\ c_{s}^{\frac{1+\beta }{%
1-\beta }}\beta \left( {\frac{k}{k_{1}}}\right) ^{-{\frac{2}{\beta -1}}}.
\label{sp}
\end{equation}
For the tensor modes, we have of course the usual result. The ``consistency
relation'' in this case reads as 
\begin{equation}
{\frac{P^{h}}{P^{\zeta }}}=-8\ c_{s}^{1-n_{T}}\frac{n_{T}}{1-n_{T}/2}.
\label{tp}
\end{equation}
Scalar and tensor perturbations have the same power dependence in $k$ and
the spectral index $n_{T}=-2/\left( \beta -1\right) $ $\ $doesn't depend on
the wavelength. Note that their relative normalization depends on the speed
of sound, and that (\ref{sp}) and (\ref{tp}) reduce to the slow roll results
in the limit of large $\beta $ (or small $n_{T}$).

{\it Discussion. } In this paper we have developed the quantum theory of
cosmological perturbations for the case of scalar field with a Lagrangian
which is an arbitrary local function of scalar field and its first
derivatives. The formalism has been derived in sufficient generality to
cover flat as well as closed and open universes. The theory gives a unified
treatment of some of the previously studied cases, which are reformulated in
a language somewhat closer to hydrodynamics, covering also the new important
case of $k-$inflation. This type of inflation is driven by non-minimal
kinetic terms, whose origin can be motivated by considering the low energy
effective action of string theory.

As an application, we have found general expressions for the power spectrum
in the case of slow roll and power law {\it k}-inflation. In particular, we
find that the ``consistency relation'' between the tensor spectral index and
the relative amplitude of tensor and scalar perturbations in {\it k-}%
inflation is different from the usual one in standard inflationary models.
Thus, in principle, {\it k-}inflation is phenomenologically distinguishable
from the usual potential-driven inflation. Also, as it was mentioned above,
the speed of sound $c_{s}$ in power law inflation can be adjusted to any
given value. This opens a new interesting possibility. Namely, one can build
an inflationary model where the tensor modes dominate over the scalar modes,
a prediction which can be confirmed or ruled out in future observations.
However, as we see from (\ref{tp}), this is only possible when the speed of
sound $c_{S}$ exceeds the speed of light. Since this is a very interesting
and rather exotic possibility, we would like to conclude with some
speculative remarks on this topic.

First of all, let us note that the theory we consider is perfectly Lorenz
invariant, but strongly nonlinear. In this case, it has been known for a
long time that the propagation of signals with the speed bigger than the
speed of light is rather generic than exceptional and does not contradict to
general principles \cite{rus} (incidentally, one expects that this
phenomenon should also occur in string theories, which predict the
Born-Infeld type action.) Returning now to the model of power law inflation,
we see that a Lagrangian of the form $p=p(X)\varphi ^{-2}$ can drive power
law expansion with an arbitrary equation of state. In particular, the scalar
field could play the role of a quintessential component in the universe with 
$p<0$. In this case too the parameters can be arranged so that $c_{s}>1$.
Therefore it doesn't seem to be excluded that we may one day be able to
communicate with distant civilizations in an efficient manner by exchanging
waves of quintessence. These speculative ideas clearly require further
investigation \cite{alex}.

{\em Acknowledgments} J.G. acknowledges support from the program
Sonderfoschungsbereichs 375, and form CICYT under contract AEN98-1093.

\end{document}